# DC THERMAL MODELING OF CNTFETS BASED ON A SEMI-EMPIRICAL APPROACH

Roberto Marani and Anna Gina Perri

Electronic Devices Laboratory, Electrical and Information Engineering Department,
Polytechnic University of Bari, via E. Orabona 4, Bari – Italy
annagina.perri@poliba.it

*ABSTRACT*

*A new DC thermal model of Carbon Nanotube Field Effect Transistors (CNTFETs) is proposed. The model is based on a number of fitting parameters depending on bias conditions by third order polynomials. The model includes three thermal parameters describing CNTFET behaviour in terms of saturation drain current, threshold voltage and M exponent in the knee region versus the temperature. To confirm the validity of the proposed thermal model, the simulations were performed in very different thermal conditions, obtaining I-V characteristics perfectly coincident with those of other models. The very low CPU calculation time makes the proposed model particularly suitable to be implemented in CAD applications.*

## 1. INTRODUCTION

Carbon NanoTubes have attracted much interest in the scientific world due to their excellent electrical, thermal, and mechanical properties. Since 1998 there have been huge developments in this new technology. In particular CNTFETs, which are field-effect transistor that utilize a single carbon nanotube or an array of carbon nanotube as the channel material instead of bulk silicon in the traditional MOSFET structure, are now a promise for an alternative material to replace silicon in future electronics [1].

Prediction through modelling forms the basis of engineering design. The computational power at the fingertips of the professional engineer is increasing enormously and techniques for computer simulation are changing rapidly. Engineers need models which relate to their design area and are adaptable to new design concepts. They also need efficient and friendly ways of presenting, viewing and transmitting the data associated with their models.

Most of the CNTFET models available in literature are numerical and make use of self-consistency and therefore they cannot be directly implemented in modeling languages for analog and digital circuits, such as SPICE, Verilog or VHDL-AMS. Moreover the research on CNTFETs thermal modeling is still at an early stage.

Generally there are two methods for developing thermal models of electronic devices [2].

The first method consists of the calculations of the thermal gradient inside the device and drain-source current $I_{ds}$, since the thermal field depends on the consumed power. This procedure allows good agreement between the measured and simulated I-V static characteristics at different temperatures but it results rather heavy from a computational point of view.

On the other hand, physical models of the temperature dependence of device parameters must take into account a number of physical factors, such as, for example, the energy gap, the available state density at the conducting and valence band edge, the dielectric constant, the electron saturation velocity, the electron mobility and the Schottky barrier height [2]. Therefore the approach generally used for the thermal modeling lies in accounting for the device parameter dependence as linear or non linear functions of temperature. In these functions temperature coefficients, depending on the device technology and bias voltages, are used. Physical modeling of bias dependence on the device characteristics can be accurate enough if analytic expressions take into account a lot of second order effects, such as the negative output conductance, especially due to self-heating and substrate currents, backgating and surface states. Thus model expressions become very complicated. This makes it very difficult to model the large signal behaviour by means of physical expressions and to derive an analytic small signal model from the large signal one. On the other hand, physical models do not required an electrical characterization of the devices.

The second method is based on a semi-empirical approach, where some of the fitting parameters of the large signal model are dependent on the temperature according to empirical relationships including fitting parameters having a physical meaning. However the semi-empirical model requires the electrical characterisation of devices and the delicate procedure of parameters extraction.

In this paper we have followed this second approach, because it can be considered more useful for CAD applications, since it is less complicated without any loss of accuracy, can be easily implemented in an electrical simulator and the computational time is very low.

The presentation is organized as follows. Section 2 gives a brief description of CNTFETs. The proposed model is presented in Section 3, while the discussion of simulation results are given in section 4. The conclusions and future developments are described in Section 5.

## 2. CARBON NANOTUBE FIELD EFFECT TRANSISTORS

The scaling of electronic devices, particularly of silicon-based transistors, has always had a key role in electronic evolution.

Today IBM has production lines with 45 nm thick channels [3], but the "International Technology Roadmap for Semiconductors" plans for less than 6 nm tick channels in 2020 [4].

The smaller and smaller scaling approaches the time when the diffusion regions of transistors will be so close that the channel will be few atoms thick and gate oxide so thin that the charge will tunnel through it: this will be the ultimate size and performance of silicon-based devices.

The Carbon NanoTubes (CNTs) have been identified as an excellent choice for next generation of field-effect transistors, which maintain the operating principles of the currently used devices, but replace the conducting channel with carbon nanomaterials such as one-dimensional (1D) CNT or two-dimensional (2D) graphene layers.

These new devices have molecular building block not coming from lithography and, along with these devices, molecular electronics will change the equation in our tool box, we will drop out well known partial differential equation for charge diffusion and we will use quantum mechanic to describe electrons, holes, atoms, molecules and photons. In coming years we will gain new tools from chemistry and physics, new sophisticated mathematical tool to face probability amplitude waves.

Carbon NanoTube Field Effect Transistors (CNTFETs), as already said, are FETs using a carbon nanotube as channel, and are regarded as an important contending device to replace conventional silicon transistors [1]. As it is known, the carbon nanotubes consist in a hexagonal mesh of carbon atoms wrapped in cylinder shapes, some time with closing hemispherical meshes on the tips. Since they could be extended for several millimeters, they have a huge length/diameter ratio making them unidimensional structures (1-D). Moreover an important characteristic of CNT is mesh torsion, denoted as *chirality*, which has a strong influence on the CNT behaviour. Depending on their torsion, electronic band structure changes, band gap can appear making them semiconductors, or cannot appear, making them conductors. Furthermore the CNT behaviour as semiconductor has an energy gap inversely proportional to their radius.

In particular we have considered the conventional CNTFETs, denoted as C-CNTFETs, with heavily doped source and drain contacts, because these devices show the best performances in terms of "on-off" ratio currents and sub-threshold swing.

# 3. THE PROPOSED DC THERMAL MODEL OF CNTFETS

Generally the empirical DC models of electronic devices for CAD purposes are developed by using mathematical functions, whose graphical representation is similar to the measured I-V curves. The use of empirical parameters enables the modeled curves to be matched to the measured curves as well as possible. The absolute error between measured and calculated current values is minimised and the parameter extraction procedure is performed. In terms of computability an empirical model is much more easily tailored to fast convergence performance, computing time and accuracy as well as second-order effects.

On the other hand, this kind of model is less intuitive than a physical model.

Consequently, much more assistance is needed for the parameter extraction of the model, especially to perform an initial estimation of the empirical parameters that will give low CPU time consumption and the extraction of the best set of parameters, allowing the best fit between measured and modeled current values.

Moreover the drain-source current is modeled as a function of internal voltages, considering the voltage drop due to the source and drain parasitic resistances. This is a limitation of this type of model, since the resistances depend on the bias conditions. In fact the parameter extraction procedure needs very difficult measurement of parasitic resistances under many bias conditions. The problem can be overcome by assuming a constant value for these resistances, but this approach affects the model with many approximations regarding the extracted parameters [5-6]. Moreover, the empirical parameters are generally considered independent of, or weakly dependent on, the bias conditions, reducing the model accuracy, particularly at the beginning of the saturation region.

The DC model, already proposed by us [7], has been implemented with the following aims:
- to improve the accuracy of modeled I-V curves, in particular in the knee and saturation regions;
- to give the device source-drain current as a function of external voltages, as seen at the device gates, by-passing the very difficult measurement of parasitic resistances for the I-V characterisation.
- to use empirical parameters to be extracted by a quick and accurate procedure, the initial estimation of empirical parameters being performed referring to measured I-V curves and to physical considerations, making univocal, fast and easy the extraction of the best fitting parameter set.

Our DC model [7], without taking into account the temperature T, has the following expression:

$$I_{ds}(V_{ds}, V_{gs}) = I_{dss}\left[\left(1-\frac{V_{gs}}{V_t}\right)^N\right][\tanh(aV_{ds})]^M \qquad (1)$$

where $I_{dss}$ is the maximum saturation drain current, which is not an empirical parameter,

$$V_t = V_{to} + \gamma * V_{ds} * V_{gs}$$

is the threshold voltage, $V_{to}$ is the threshold voltage at zero bias, $\gamma$ is the threshold voltage shift parameter, and

$$N = N_0 + N_1 V_{gs} + N_2 V_{gs}^2 + N_3 V_{gs}^3 \qquad (2a)$$

$$\alpha = a_0 + a_1 V_{gs} + a_2 V_{gs}^2 + a_3 V_{gs}^3 \qquad (2b)$$

being $V_{ds}$ the drain-source voltage and $V_{gs}$ the gate-source voltage.

Equations (2) allow the third order dependence of N and $\alpha$ on the bias conditions, improving the fitting in the linear, knee and saturation regions. Moreover, the M parameter improves the fitting in the knee region, modifying here the behaviour of hyperbolic tangent function.

The voltages in eqn. (1) are external, i.e. measured at the device external terminals. In this way it is possible to overcome the problem of measurement of the parasitic resistances, thus making easier the parameter extraction procedure and the use of the model for circuit design. If the complete DC device characterisation requires the resistances to be determined, the linear approximation can be used for them without affecting the I-V model accuracy.

The extraction of the 11 empirical parameters has been obtained by an appropriate optimisation routine, that minimizes the root-mean-squared (RMS) difference between the function and the simulation results.

The extraction of the 11 empirical parameters by an appropriate optimisation routine is widely described in our Ref. [7] and therefore the reader is requested to consult it.

In this paper we modify the previous DC model to simulate very accurately the effect of device negative output conductance, especially due to self-heating and substrate currents.

We propose that thermal effects can be modelled by the dependence on temperature T of the device threshold voltage and of the maximum saturation drain-source current, because of their strong sensitivity to the temperature variations as suggested by the device physical behaviour analysis. Besides, in linear region the current $I_{ds}$ increases as the temperature increases. In the saturation region the drain current decreases with temperature. This behaviour is shown in $I_{ds}$ - $V_{ds}$ curves of FET with a long gate length. In short gate devices thermal effects are evident only in saturation region. Moreover, internal thermal effects cause mainly a decrease of current in saturation region, that is the output negative differential conductance phenomenon for high gate-source voltage $V_{gs}$, occurring as a consequence of the reduced mobility and saturation velocity. In correspondence to low $V_{gs}$, near the threshold voltage, there is an increase of current with temperature. This phenomenon can be modeled as a shift in the threshold voltage, that decreases as the temperature increases.

Therefore the proposed thermal DC model has the following expression is:

$$I_{ds}(V_{ds}, V_{gs}, T) = I_{dss}(T)\left[\left(1 - \frac{V_{gs}}{V_t(T)}\right)^N - k(1 + \lambda V_{ds})e^{-cV_{gs}^2}\right][\tanh(\alpha V_{ds})]^M \quad (3)$$

The dependence of $I_{dss}$ on temperature is modeled as:

$$I_{dss}(T) = I_{dss}(T_o) \times \exp[-k_1 * (T - T_o)] \quad (4)$$

where $T_o$ is the room temperature and $I_{dss}(T_o)$ has a known measured value.

The dependence of $V_t$ on temperature is modeled as:

$$V_t(T) = V_{to}(T_o) - k_2 * (T - T_o) + \gamma * V_{ds} * V_{gs} \quad (5)$$

To improve the model predictions in the knee region, it is appropriate to consider also the M parameter as a linear function of T, as:

$$M(T) = M(T_o) - k_3(T - T_o)*(1 + \eta V_{ds}) \quad (6)$$

being typically $\eta = 1$ V$^{-1}$ and therefore it must not be considered a parameter to be determined.

The presence of drain-source voltage $V_{ds}$ in Eqn.(6) allows to make accurate predictions of the cross point among curves at different temperatures, i.e. to predict the position of the insensitive-temperature points. The variation of M due to thermal effects is particularly significant in FETs having a long channel, where the current $I_{ds}$ in the linear region is strongly dependent on the temperature.

The empirical parameters are extracted with high accuracy in three steps.

The first two steps are executed on data measured at room temperature, according to the procedure described in Ref. [7]. At first, the $V_{to}$, $\gamma$ and $N_i$ coefficients are extracted by considering the I-V measurements only in the saturation region. The determination of k, $\lambda$ and c parameters does not require an extraction procedure.

As a second step, the parameters M ($T_o$) and $a_i$ (coefficients of $\alpha$), characterizing the device behaviour in the linear region, are calculated minimizing the error between measured and modeled current values in all field of measurements, i.e. in the linear and saturation region.

Finally, the third step in the procedure extraction consists of the determination of the thermal fitting parameters, i.e. $k_1$, $k_2$ and $k_3$, by using the data measured at different temperatures. Their initial values are generally assumed very low, around $10^{-4}$. Therefore, the maximum total number of empirical fitting parameters is 14.

It is not a quite high number, as it could seem, because the fitting parameters are not always all present and the parameters are nor extracted all at the same time.

## 4. SIMULATION RESULTS AND DISCUSSION

Firstly, we have extracted the fitting parameters, i.e. $V_{to}$, $\gamma$, $N_o$, $N_1$, $N_2$, $N_3$, $a_o$, $a_1$, $a_2$, $a_3$, M by minimizing a squared error function within the required tolerance (typically $10^{-4}$), at room temperature. This fitting parameter extraction was performed in two steps: in the first, only the parameters $V_{to}$, $\gamma$, $N_o$, $N_1$, $N_2$, $N_3$ involved in saturation region are extracted. In the second step, the remaining parameters characterizing the linear region, i.e. $a_o$, $a_1$, $a_2$, $a_3$, M, are obtained.

The initial parameter estimation and the resulting extracted parameters to fit the measured characteristics [8-9] are summarised in Table 1.

Table 1. Results of extraction procedure at room temperature.

| Fitting parameters | Estimated parameters | Extracted parameters |
|---|---|---|
| $V_{to}$ (V) | -3.8 | -2.63 |
| $\gamma$ ($V^{-1}$) | 0.03 | 0.0377 |
| $N_o$ | 2 | 1.2669 |
| $N_1$ ($V^{-1}$) | 0 | $1.1092 \cdot 10^{-4}$ |
| $N_2$ ($V^{-2}$) | 0 | $8.9481 \cdot 10^{-5}$ |
| $N_3$ ($V^{-3}$) | 0 | $1.1988 \cdot 10^{-4}$ |
| $a_o$ ($V^{-1}$) | 3 | 1.5314 |
| $a_1$ ($V^{-2}$) | 0 | -0.2382 |
| $a_2$ ($V^{-3}$) | 0 | 0.1784 |
| $a_3$ ($V^{-4}$) | 0 | 0.0279 |
| M | 1 | 1.3779 |

The strong improvement of accuracy is fundamentally due to the third-order dependence of $\alpha$ and N on voltage $V_{gs}$ and to the dependence of $V_t$ on $V_{gs}$. Moreover, the model parameter extraction procedure is fairly quick and unique due to the close estimation of their initial values to start the extraction routine. Finally, the CPU execution times for the model parameter extraction and for the model calculation are reasonably small.

Then we have plotted the output and transfer curves for a C-CNTFET, whose technological characteristics are reported in [10].

Fig. 1 shows the $I_{DS} - V_{DS}$ characteristics (denoted by continuous lines) of numerical simulations, according to our procedure, and the experimental ones [8-9] (denoted by ●).

As shown in Fig.1, we have obtained a very good agreement between measured and modelled data in terms of $I_{ds}$ versus $V_{ds}$, with a relative error less than 4 %.

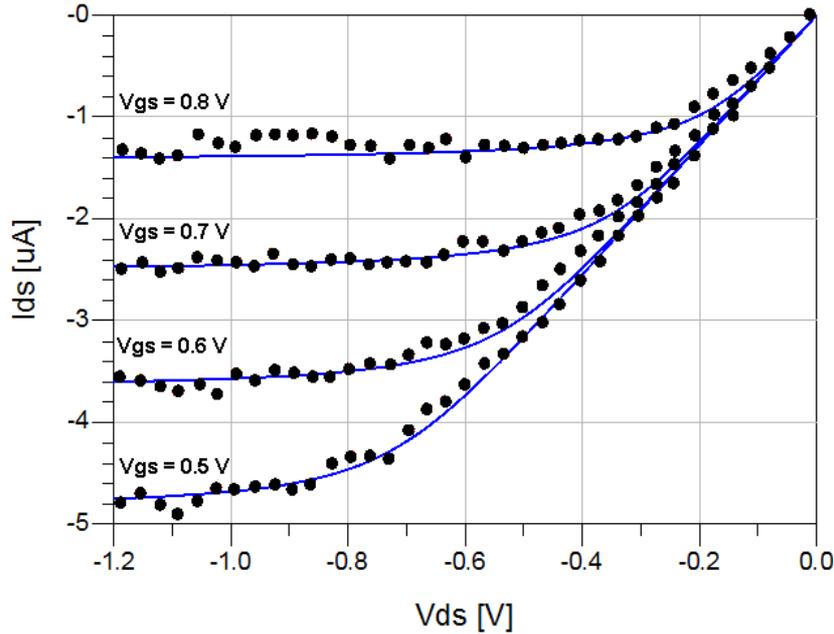

Figure 1. Simulated $I_{DS} - V_{DS}$ characteristics (denoted by continuous lines) and experimental $I_{DS} - V_{DS}$ characteristics [8-9] (denoted by ●).

Moreover the CPU calculation time was very short, 0.03 s, by using a 2.4 GHz compatible PC.
We have also obtained a good agreement between measured and simulated transfer characteristics [7] and our results have been also compared with those of the numerical model, named FETToy, online available [11], obtaining a negligible relative error also in case (less than 5%), with a CPU calculation time much more low.
Now, in order to confirm the validity and to verify the accuracy of the proposed thermal model, we have compared our results with those of CNTFET model proposed in [12].
At first we have extracted the parameters depending on thermal effects, i.e. $V_{to}$ and $I_{dss}$, evaluating the root mean-square errors between our thermal model and the numerical ones shown in [12].
The simulations were performed in very different thermal conditions (i.e. at temperatures of 50 K, 300 K and 400K) and the results, reported in Table 2 only for CNTFET of [12] having a diameter of 1 nm, show that the calculation error is very low, |E| = 1.71 mA, with relative percentage of 4.4%.

Table 2. Extracted parameters depending on thermal effects at different temperatures.

| T (K) | $V_{to}$ (V) | $I_{dss}$ (mA) | \|E\| (mA) | E (%) |
|---|---|---|---|---|
| 50 | -0.968 | 37.79 | 0.07 | 1.7 |
| 300 | -2.100 | 61.74 | 1.71 | 4.4 |
| 400 | -2.240 | 54.57 | 1.27 | 2.3 |

Moreover the CPU calculation time was very short, of about 0.01 s, by using a 2.4 GHz compatible PC. This result allows us to assert that our model is suitable to be implemented in CAD tools.
Regards the I-V characteristics, our results are perfectly coincident with those obtained by the model proposed in [12-13].
In Fig. 2 we have reported only one curve in order to clearly see the effects of temperature. In particular curves 1, 2 and 3 indicate our simulated I-V characteristics at temperatures of 50 K, 300 K and 400 K respectively.

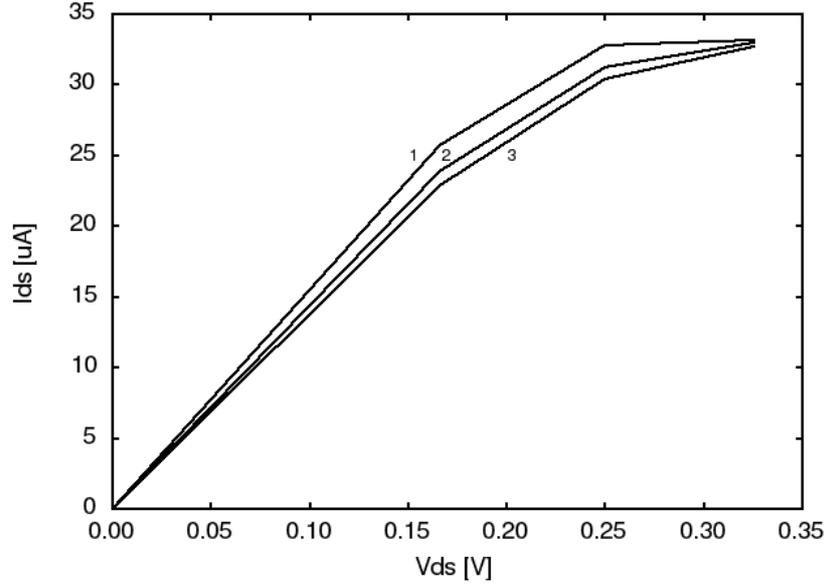

Figure 2. Simulated $I_{DS} - V_{DS}$ characteristics at different temperatures for CNTFET having a diameter of 1 nm.

The maximum current is affected as temperature increases and the linear region seems to not be affected much by temperature from $V_{DS}$ voltages of 0 V to 0.2 V.

Moreover the proposed model is directly identified without prior extraction of the series parasitic resistances and this clearly simplifies the procedure for model extraction from measured data, but this is a limitation of this type of model, since the resistances depend on the bias conditions. In fact the parameter extraction procedure needs very difficult measurement of parasitic resistances under many bias conditions. The problem can be overcome by assuming a constant value for these resistances, but this approach affects the model with many approximations regarding the extracted parameters.

However, in [5-7] and in [14-15], we have investigated about the effects of the CNT parasitic resistances, demonstrating that they can be neglected, for both analog and digital applications, at frequencies under ten THz.

## 5. CONCLUSION AND FUTURE DEVELOPMENTS

In this paper we have proposed a semi-empirical large signal thermal model of CNTFETs. The model is based on a number of fitting parameters depending on bias conditions by third order polynomials. The model includes three thermal parameters describing the CNTFET behaviour in terms of saturation drain current, threshold voltage and M exponent in the knee region versus the temperature.

The main aims have been to improve the accuracy of modelled I-V curves, in particular in the knee and saturation regions; to give the device source-drain current as a function of external voltages, as seen at the device gates, by-passing the very difficult measurement of parasitic resistances for the I-V characterisation and to use empirical parameters to be extracted by a quick and accurate procedure, the initial estimation of empirical parameters being performed referring to measured I-V curves and to physical considerations, making univocal, fast and easy the extraction of the best fitting parameter set. The extraction of the empirical parameters of the model has been obtained by an appropriate optimisation routine, that minimizes the root-mean-squared difference between the function and the simulation results.

The proposed model has been initially utilized to reproduce the measured I - V characteristic curves at room temperature, obtaining a very good agreement both between measured [8-9] and modelled data and between our data and those obtained from simulator online available [11], with a CPU calculation time much low.

Then we have modified our DC model to simulate very accurately thermal effects, especially due to self-heating and substrate currents, introducing the dependence on temperature of the device threshold voltage and of the maximum saturation drain-source current

The simulations were performed in very different thermal conditions and our results have been compared with those of CNTFET model proposed in [12], obtaining, with reference to I-V characteristics, trends perfectly coincident, but with CPU calculation times much lower. This result allows us to assert that the proposed model seems particularly suitable to be implemented in CAD applications.

Actually we are implementing a measurement apparatus to characterize in temperature the CNTFETs behaviour.

# Authors

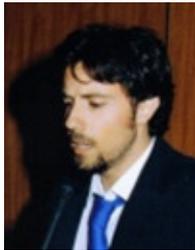

**Roberto Marani** received the Master of Science degree (*cum laude*) in Electronic Engineering in 2008 from Polytechnic University of Bari, where he received his Ph.D. degree in Electronic Engineering in 2012.

He worked in the Electronic Device Laboratory of Bari Polytechnic for the design, realization and testing of nanometrical electronic systems, quantum devices and FET on carbon nanotube. Moreover Dr. Marani worked in the field of design, modelling and experimental characterization of devices and systems for biomedical applications.

In December 2008 he received a research grant by Polytechnic University of Bari for his research activity. From February 2011 to October 2011 he went to Madrid, Spain, joining the Nanophotonics Group at Universidad Autónoma de Madrid, under the supervision of Prof. García-Vidal.

Currently he is involved in the development of novel numerical models to study the physical effects that occur in the interaction of electromagnetic waves with periodic nanostructures, both metal and dielectric. His research activities also include biosensing and photovoltaic applications.

Dr. Marani is a member of the COST Action MP0702 - Towards Functional Sub-Wavelength Photonic Structures, and is a member of the Consortium of University CNIT – Consorzio Nazionale Interuniversitario per le Telecomunicazioni.

Dr. Marani has published over 100 scientific papers.

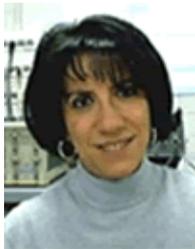

**Anna Gina Perri** received the Laurea degree *cum laude* in Electrical Engineering from the University of Bari in 1977. In the same year she joined the Electrical and Electronic Department, Polytechnic University of Bari, Italy, where she is Full Professor of Electronics from 2002.

From 2003 she has been associated with the National Institute of Nuclear Phisics (INFN) of Napoli (Italy), being a part of the TEGAF project: "Teorie Esotiche per Guidare ed Accelerare Fasci", dealing with the optimal design of resonance-accelerating cavities having very high potentials for cancer hadrontherapy.

In 2004 she was awarded the "Attestato di Merito" by ASSIPE (ASSociazione Italiana per la Progettazione Elettronica), Milano, BIAS'04, for her studies on electronic systems for domiciliary teleassistance.

Her current research activities are in the area of numerical modelling and performance simulation techniques of electronic devices for the design of GaAs Integrated Circuits and in the characterization and design of optoelectronic devices on PBG (Phothonic BandGap).

Moreover she works in the design, realization and testing of nanometrical electronic systems, quantum devices, FET on carbon nanotube and in the field of experimental characterization of electronic systems for biomedical applications.

Prof. Perri is the Head of the Electron Devices Laboratory of the Polytechnic University of Bari.

She has been listed in the following volumes: Who's Who in the World and Who's Who in Engineering, published by Marquis Publ. (U.S.A.).

She is author of over 250 journal articles, conference presentations, twelve books and currently serves as a Referee of a number of international journals.

Prof. Perri is the holder of two italian patents and the Editor of two international books.

She is also responsible for research projects, sponsored by the Italian Government.

Prof. Perri is a member of the Italian Circuits, Components and Electronic Technologies – Microelectronics Association, and an Associate Member of National University Consortium for Telecommunications (CNIT).